\documentclass[conference]{IEEEtran}
\IEEEoverridecommandlockouts
\usepackage{cite}
\usepackage{amsmath,amssymb,amsfonts, bbm}
\usepackage{algorithm, algorithmic}
\usepackage{subcaption}
\usepackage{graphicx}
\usepackage{textcomp}
\usepackage{multirow}
\usepackage{longtable}
\usepackage{multicol}
\usepackage{float}
\usepackage{url}
\usepackage{hyperref}

\usepackage{listings}
\usepackage{color}

\definecolor{dkgreen}{rgb}{0,0.6,0}
\definecolor{gray}{rgb}{0.5,0.5,0.5}
\definecolor{mauve}{rgb}{0.58,0,0.82}

\usepackage{xcolor}
\usepackage{placeins} 
\def\BibTeX{{\rm B\kern-.05em{\sc i\kern-.025em b}\kern-.08em
    T\kern-.1667em\lower.7ex\hbox{E}\kern-.125emX}}

\begin{document}

\title{Chronological Outlooks of Globe Illustrated with Web-Based Visualization}
\author{\IEEEauthorblockN{Tahmim Hossain}
\IEEEauthorblockA{\textit{Department of Computing Science} \\
\textit{University of Alberta}\\
Edmonton, Canada \\
tahmim@ualberta.ca}
\and
\IEEEauthorblockN{Sai Sarath Movva}
\IEEEauthorblockA{\textit{Department of Computing Science} \\
\textit{University of Alberta}\\
Edmonton, Canada \\
saisarat@ualberta.ca}
\and
\IEEEauthorblockN{Ritika Ritika}
\IEEEauthorblockA{\textit{Department of Computing Science} \\
\textit{University of Alberta}\\
Edmonton, Canada \\
ritika7@ualberta.ca}
}

\maketitle

\begin{abstract}

Developing visualizations with comprehensive annotations is crucial for research and educational purposes. We've been experimenting with various visualization tools like Plotly, Plotly.js, and D3.js to analyze global trends, focusing on areas such as Global Terrorism, the Global Air Quality Index (AQI), and Global Population dynamics. These visualizations help us gain insights into complex research topics, facilitating better understanding and analysis. We've created a single web homepage that links to three distinct visualization web pages, each exploring specific topics in depth. These webpages have been deployed on free cloud hosting servers such as Vercel and Render.

\end{abstract}


\section{Introduction}

\noindent Visualizations are essential for translating complex data into understandable insights in today's data-rich environment. They offer a powerful tool for dissecting global trends, such as terrorism, air quality, and population dynamics. By analyzing data through interactive maps, timelines, and charts, viewers can explore and understand multifaceted phenomena. These visualizations not only highlight areas of concern but also empower policymakers and researchers to devise more effective strategies for prevention and response.

\noindent For instance, visualizing global terrorism trends reveals insights into the spatial dynamics of terrorist activity, including hotspots, patterns of escalation, and underlying causes. Similarly, visualizations of the Global Air Quality Index provide real-time insights into air pollution levels worldwide, aiding in understanding health risks and prompting action towards cleaner environments. Additionally, visualizations of global population dynamics illustrate demographic transitions, informing decisions on resource allocation and social welfare programs.
 
 \noindent Paper \cite{r1} delves into geographical perspectives on terrorism, showcasing its role in unraveling the spatial dynamics of terrorist activity. Paper \cite{r2} presents an innovative air quality visualization scheme leveraging deep learning for enhanced feature extraction from multivariate time series data, aiding analysts in data exploration through rich visualizations and interactive tools. Paper \cite{r3} visualizes the global demographic transition, spotlighting the shift from rapid population growth to a new equilibrium driven by declining fertility rates and slower population growth.

\noindent In our project, we tried to utilize advanced visualization tools like Plotly, Plotly.js, and D3.js to explore these global trends. By weaving together data and visualizations, it aims to shed light on pressing issues and foster understanding among researchers and the public. Deployed on accessible web platforms, these visualizations democratize access to valuable information and promote collaboration across borders.

\section{Dataset}

\noindent We have used three datasets to visualize. The description of all three datasets has been given below:

\subsection{Global Terrorism Dataset \cite{r4}:}

\noindent The Global Terrorism CSV Database (GTD) \cite{r4} is a comprehensive dataset spanning from 1970 to 2021, containing information about terrorist incidents worldwide. Originally comprising 214,666 rows and 135 columns (sized 189 MB), after preprocessing, it consists of 214,666 rows and 20 columns (sized 23.8 MB), with null and "-99" values replaced with "0". Figure \ref{fig::1}(a) shows the preprocessed dataset.

\subsection{Global Air Quality Dataset \cite{r5}:}

\noindent The CSV dataset \cite{r5} merges city information with geographical coordinates and air pollution data across countries to offer insights into air quality, aiding policymakers and researchers in pollution mitigation efforts. Despite preprocessing, duplicate values persist. Initially containing 16,696 rows x 14 columns (sized 1.34 MB), preprocessing reduced it to 16,394 rows x 14 columns (sized 1.34 MB). Figure \ref{fig::1}(b) depicts the post-preprocessing dataset.

\subsection{Global Population Dataset \cite{r6}:}
\begin{figure}[H]
    \centering
	\includegraphics[width=0.48\textwidth]{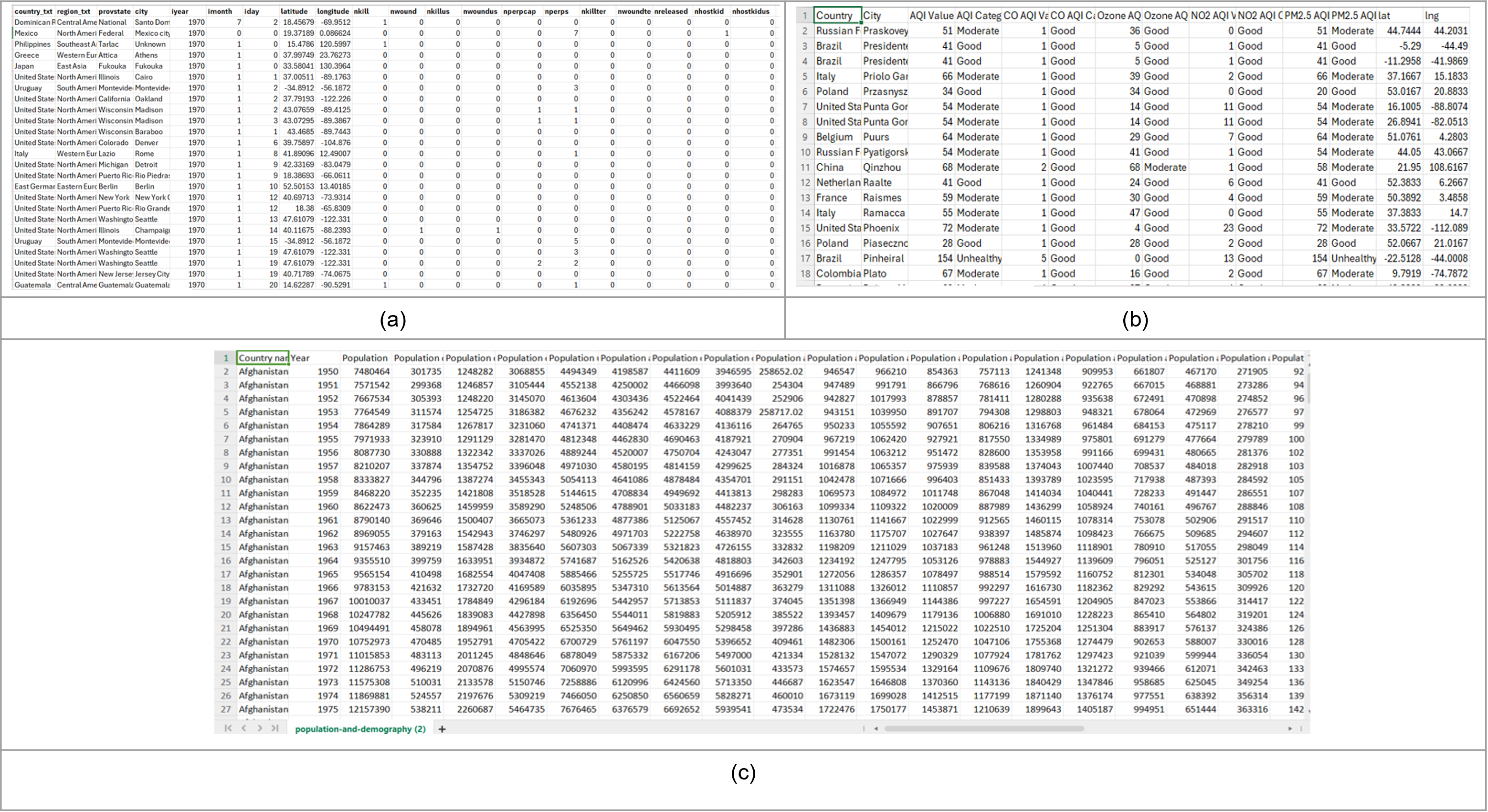}
	\caption{(a) Modified Global Terrorism Database (b) Modified Air Quality Index Dataset (c) Modified Global Population Dataset}
	\label{fig::1}
\end{figure}

\noindent The 2021 Revision of World Population Prospects CSV formatted dataset \cite{r6}, the UN's twenty-seventh edition, offers official population estimates and projections from 1950 to the present for 237 countries. The dataset was preprocessed. We didn't need to preprocess again. The dataset comprises 18289 rows and 40 columns (sized 4.54 MB). In figure \ref{fig::1}(c) shows the dataset after preprocessing.

\section{Methodology}
\begin{figure}[H]
    \centering 
	\includegraphics[width=0.4\textwidth]{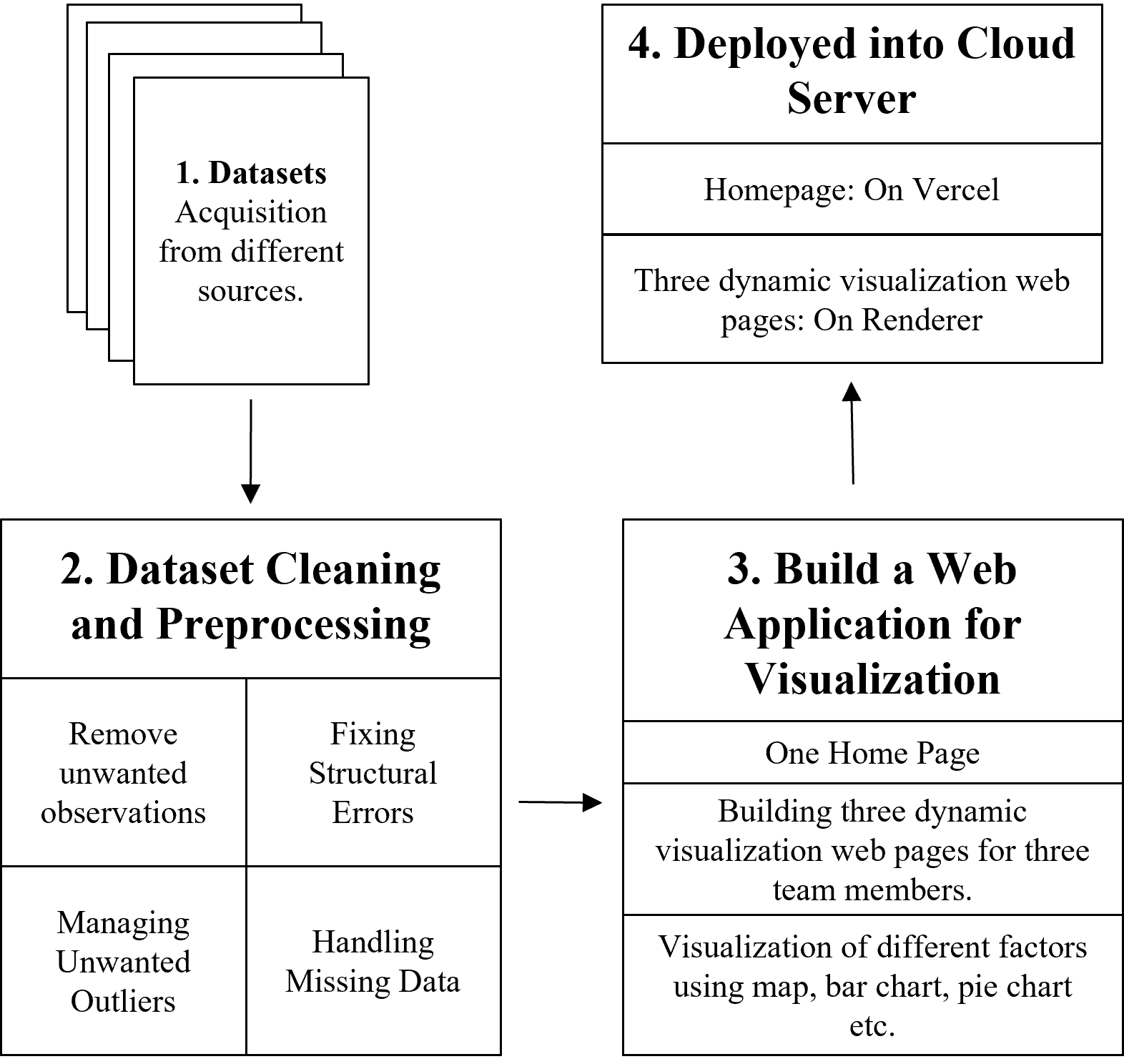}
	\caption{Methodology of our project.}
	\label{fig::4}
\end{figure}

\noindent Figure \ref{fig::4} outlines our project methodology. We started by acquiring datasets from various sources (Section II). Then, we preprocessed the data by cleaning, handling missing values, and fixing errors. Next, we developed three dynamic visualization web pages and linked them through a homepage. We deployed the homepage on Vercel and the visualizations on Renderer due to deployment limitations. Each team member deployed one visualization page. We aimed to explore different visualization libraries like Plotly Express, Plotly.js, and D3.js simultaneously, requiring separate deployments to address compatibility issues. The homepage URL is \url{https://homepage-mm804.vercel.app/}, but please note that the websites may take up to a minute to load after clicking "SEE MORE" as they are on free servers.

\section{Technology Stack}
\noindent As we built four web pages, we explored many languages, frameworks, dynamic visualization libraries, etc. The technologies that we have used are given in figure \ref{fig::11} categorizing the web pages:

\begin{figure}[H]
    \centering 
	\includegraphics[width=0.48\textwidth]{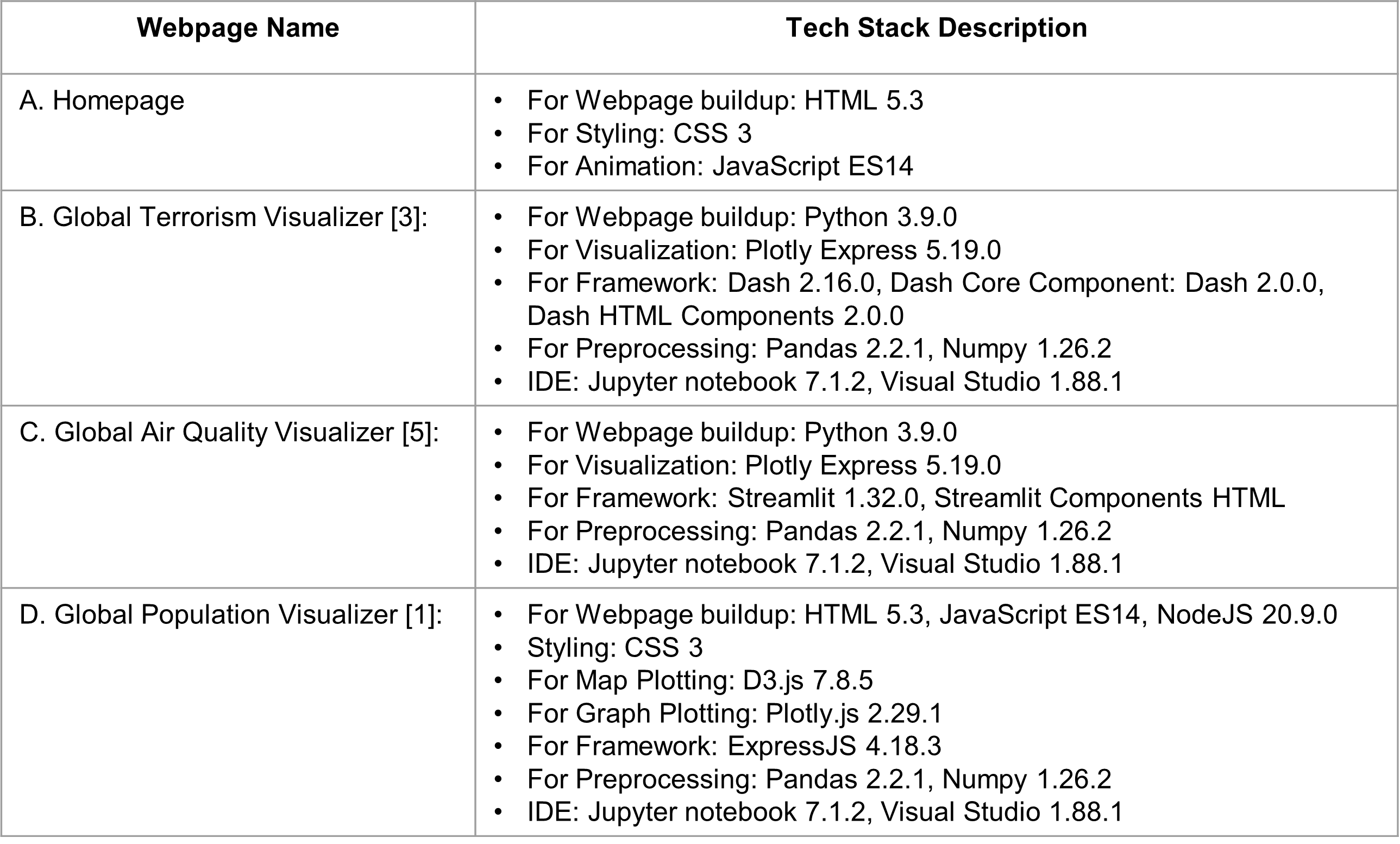}
	\caption{Technologies that were used behind the project.}
	\label{fig::11}
\end{figure}

\section{Implementation}

\noindent The implementation is discussed in this section.

\subsection{Homepage:}
\noindent We've designed a CSS file to style the webpage elements and created a JavaScript file to enable a carousel feature. This feature allows users to navigate slides using the next and previous buttons, with automatic transitions between slides. The HTML file serves as the structure for the homepage. To directly access the website go to \url{https://homepage-mm804.vercel.app/}.
\begin{figure}[H]
    \centering 
	\includegraphics[width=0.48\textwidth]{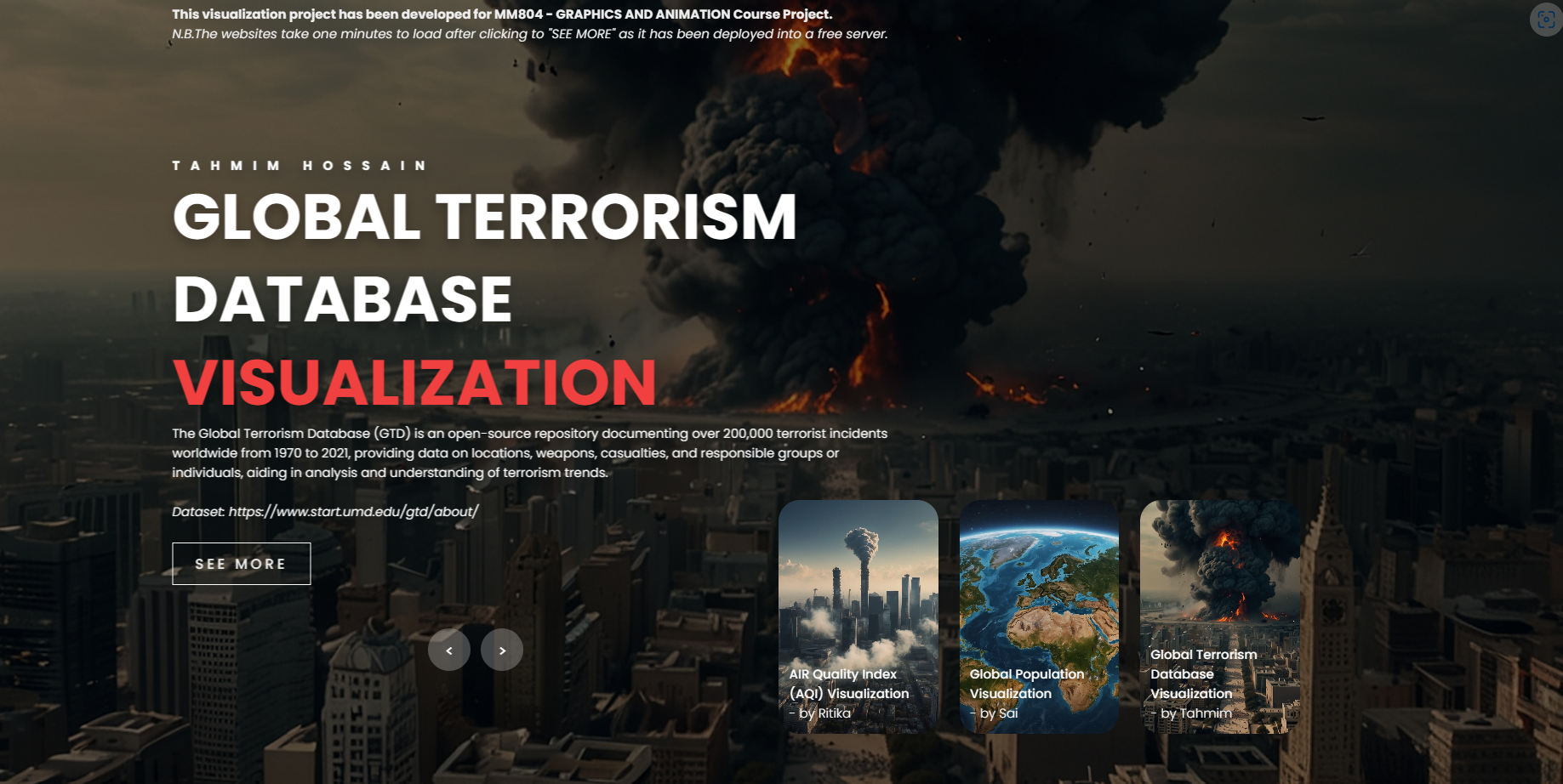}
	\caption{Homepage}
	\label{fig::5}
\end{figure}

\noindent In Figure \ref{fig::5}, the homepage is displayed, allowing users to navigate through three visualization web pages by clicking the left and right arrow signs. When users click the "SEE MORE" button, it may take a few minutes for the web page to load, as it is hosted on a free cloud server.

\subsection{Global Terrorism Visualizer \cite{r4}:}
\begin{figure}[H]
    \centering 
	\includegraphics[width=0.48\textwidth]{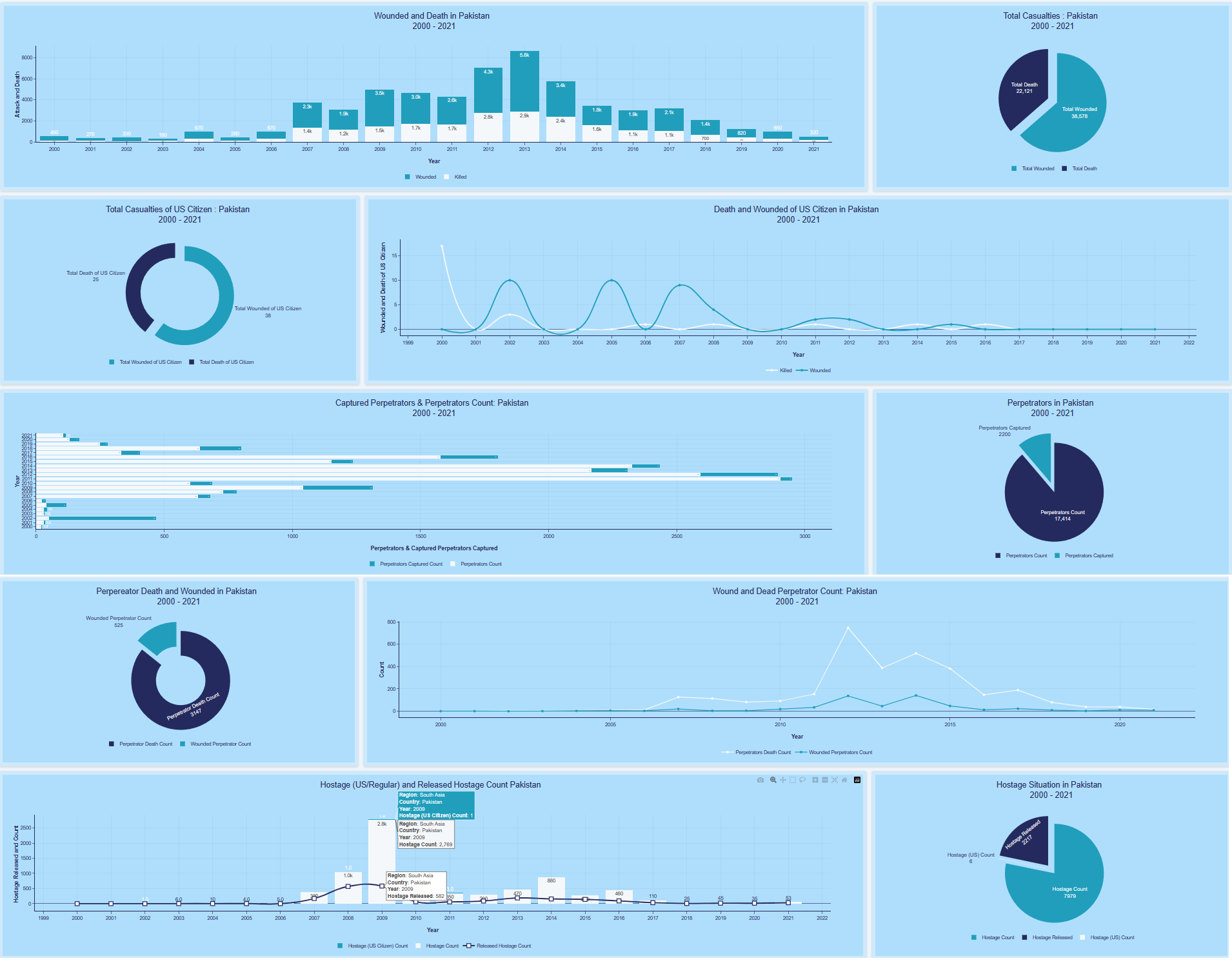}
	\caption{Hovering information inside charts for different factors.}
	\label{fig::8}
\end{figure}

\noindent Initially, we conducted data preprocessing by analyzing the dataset, addressing null and unwanted values, and selecting specific columns as outlined in the "Dataset" Section. Subsequently, we exported the modified dataset. Following this, we created two CSS files—one for webpage layout and another for styling purposes. The main Python file utilizes Plotly to generate visualizations based on user filtering. To directly access the website go to \url{https://global-terror-visual.onrender.com/}.
\begin{figure}[H]
    \centering 
	\includegraphics[width=0.48\textwidth]{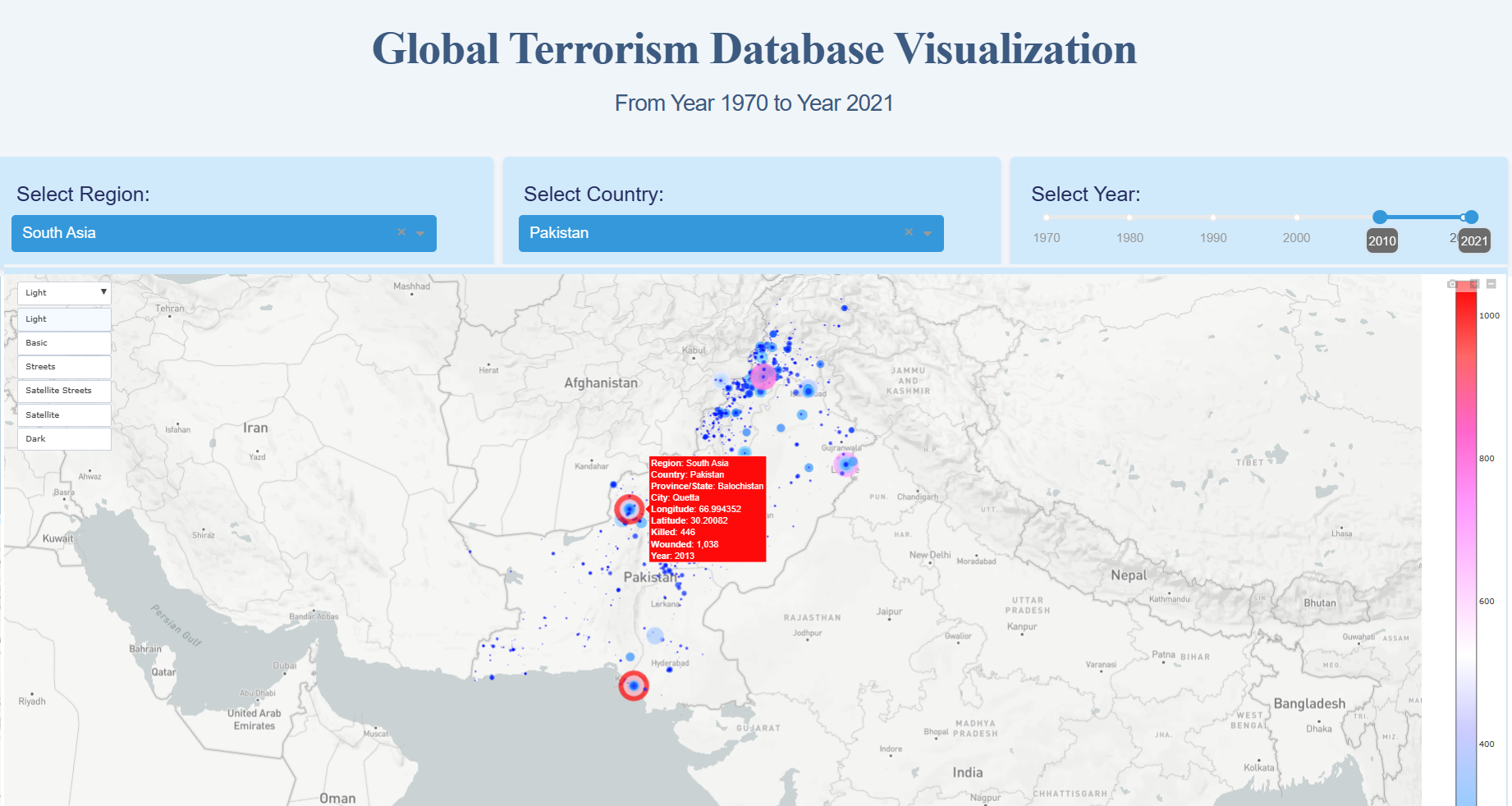}
	\caption{User Selection from Region, Country and Year and map information hovering.}
	\label{fig::6}
\end{figure}
\noindent Figure \ref{fig::6} enables users to select Region, Country, and year, updating both map and charts accordingly with bar, line, and pie charts. Map hover provides details, and users can switch between map layouts. In Figure \ref{fig::8}, users explore correlations with bar or line charts and accompanying percentages in pie charts, addressing factors like Wounded versus Death Count, Wounded US Citizens versus Death of US Citizens, Captured Perpetrators versus Perpetrators Count, Death Count of Perpetrators versus Wounded Count of Perpetrators, and Hostages as US citizens versus regular people with released hostage counts.

\subsection{Global Air Quality Visualizer \cite{r5}:}
\noindent After preprocessing the dataset as described in the "Dataset" section, we exported the modified dataset. The main Python file employs Plotly to generate visualizations based on user filtering. Streamlit is utilized for webpage styling; hence, no CSS file is required. To directly access the website go to \url{https://aqi-visualiser.onrender.com/}.
\begin{figure}[H]
    \centering 
	\includegraphics[width=0.48\textwidth]{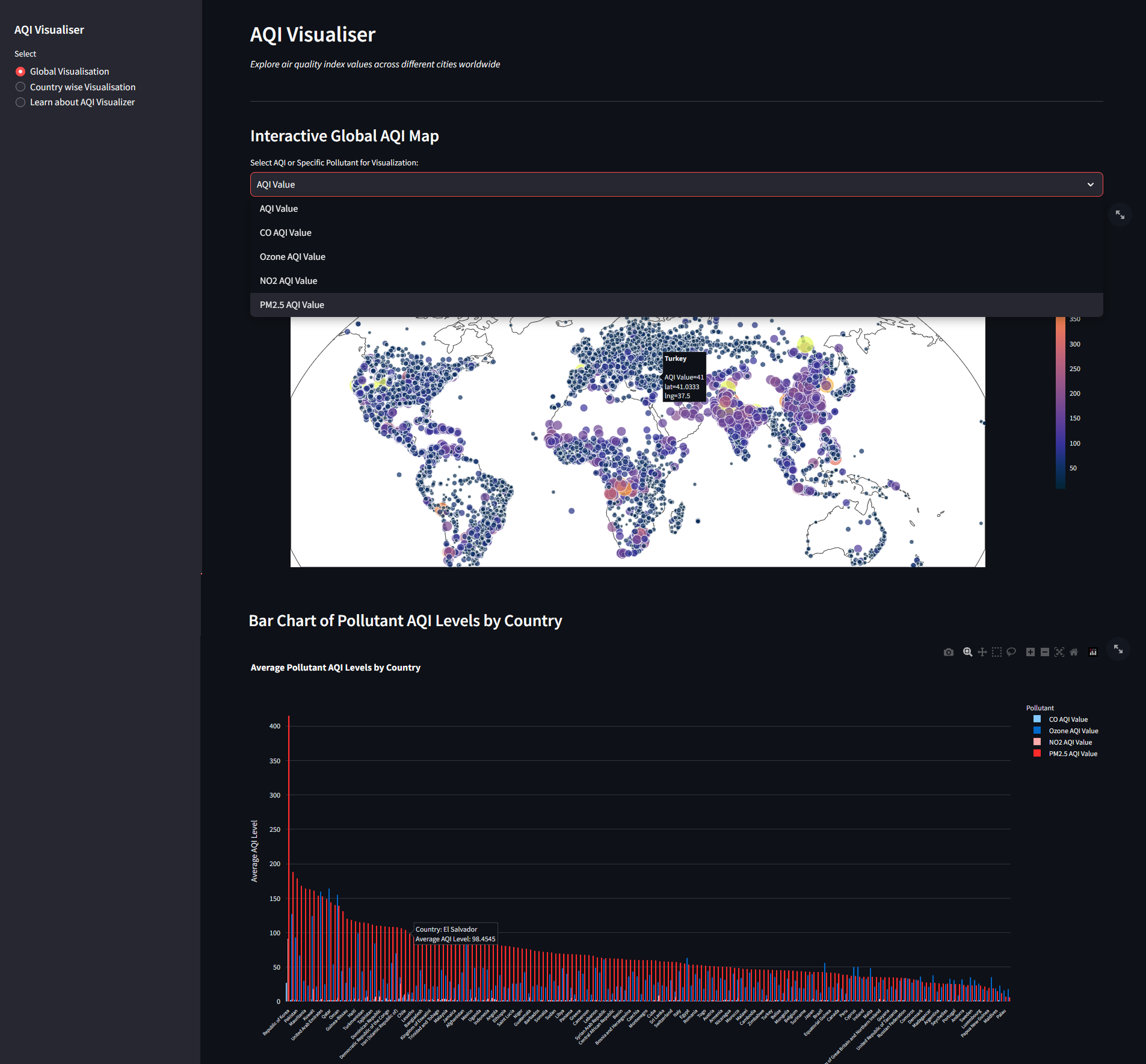}
	\caption{Air Quality Index Global Visualization}
	\label{fig::9}
\end{figure}

\begin{figure}[H]
    \centering 
	\includegraphics[width=0.48\textwidth]{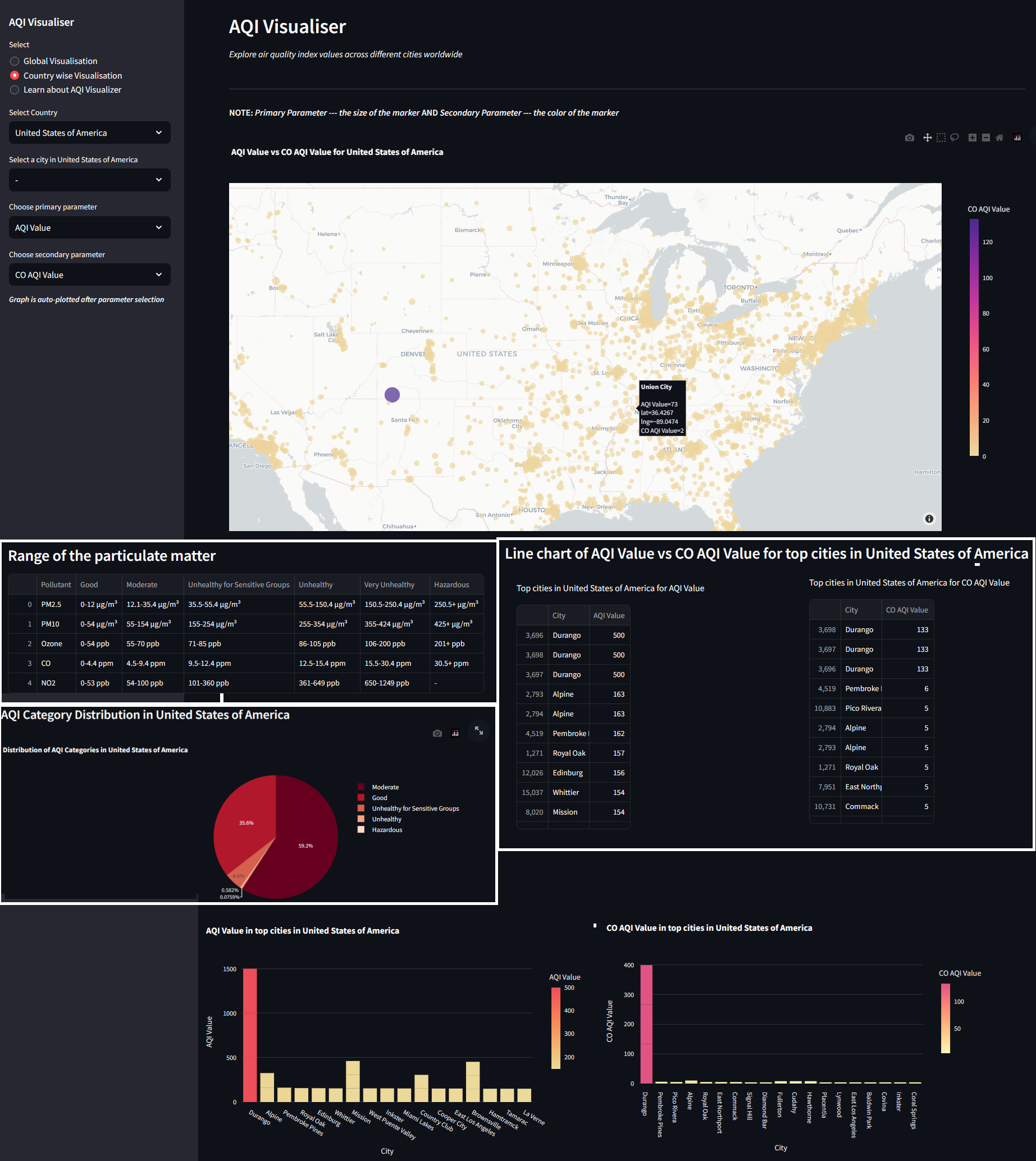}
	\caption{Air Quality Index Country-wise Visualization}
	\label{fig::10}
\end{figure}

\noindent 
Figure \ref{fig::9} offers users global or country-wise visualizations, allowing the selection of parameters like AQI, Carbon Monoxide, Ozone, Nitrogen Dioxide, and PM 2.5. They can inspect parameter values on the global map and view a bar chart of AQI levels by country. In Figure \ref{fig::10}, users choose country-wise visualization, selecting country, city, and primary/secondary parameters. The visualization includes a comparison map, top tabular data on particulate matter ranges and top cities, and bar/pie charts showing AQI distribution.

\subsection{Global Population Visualizer \cite{r6}:}
\noindent Just like the previous dataset we conducted data preprocessing by analyzing the dataset, addressing no changes for the modified dataset as it was preprocessed. Following this, two JS files have been built first one sets up a web server using Express, reads and processes data from a CSV file, and serves the data and HTML file to clients upon request and the second one fetches GeoJSON data representing countries from a URL. We used CSS inside the HTML file to style the webpage. To directly access the website go to \url{https://visualization-of-world-population.onrender.com/}.

\begin{figure}[H]
    \centering 
	\includegraphics[width=0.48\textwidth]{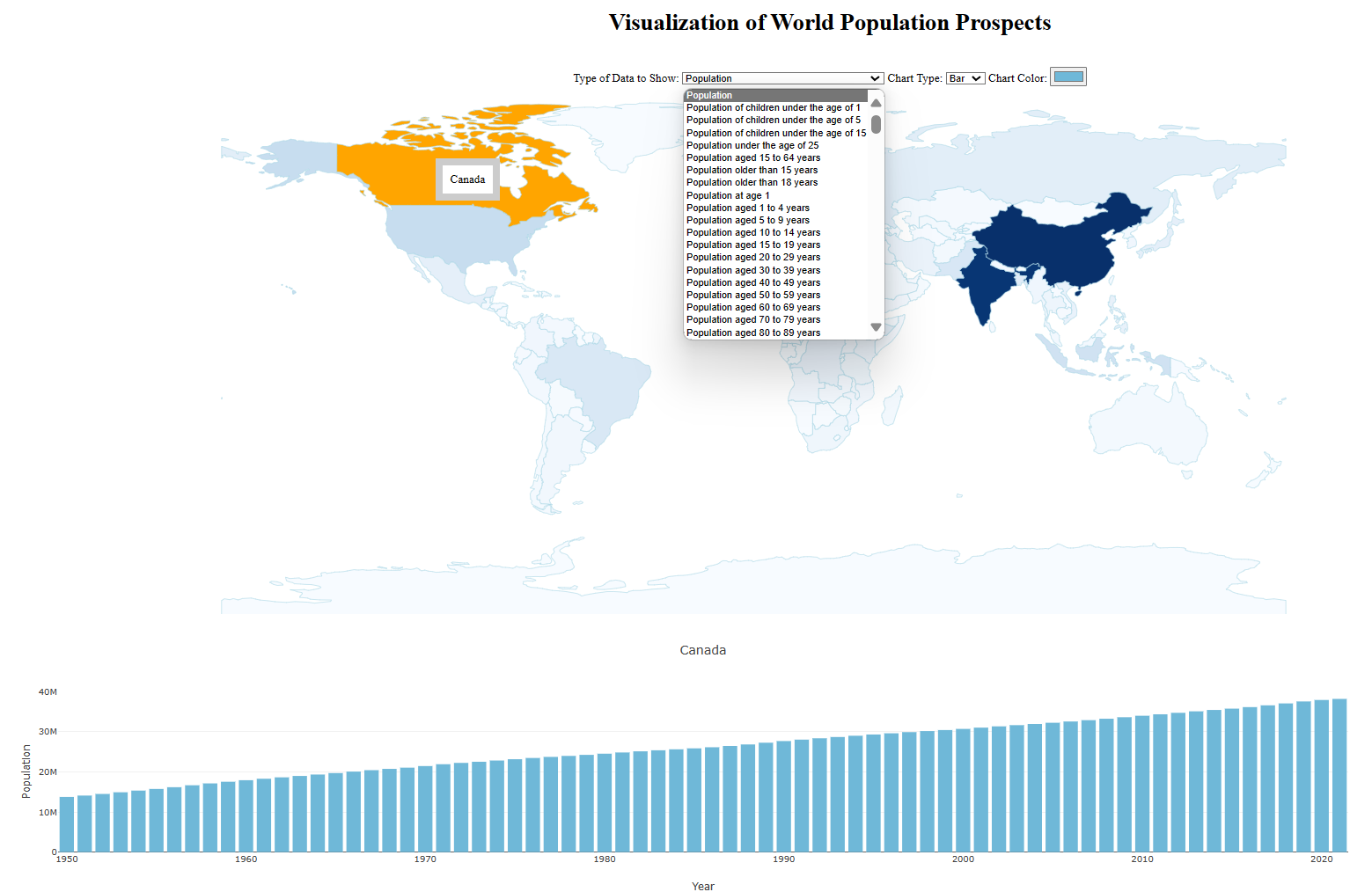}
	\caption{Global Population Dataset Visualization}
	\label{fig::12}
\end{figure}

\noindent Figure \ref{fig::12} demonstrates that users can input factors by selecting "Type of Data to show" and choosing between bar or line chart types. Additionally, users can select colors. Clicking on a specific region on the map displays the population over the years below the map.

\section{Validation Survey}

We used a Google Form \textit{\url{https://forms.gle/e9b2Fp3BHjSomnf87}} to do a validation survey for our project. We sent our website and Google form links to our random friends, and family through WhatsApp and Messenger to get anonymous feedback from them. We asked five questions and got sixteen responses. Questions were: 
\begin{enumerate}
    \item The features of web applications are easy to understand.
    \item The overall UI design of the web app.
    \item The web app allowed me to get the perfect visualization through the globe.
    \item The graphs give a perfect understanding of the behavior of different parameters.
    \item The overall satisfaction of usefulness, ease of application.
\end{enumerate}

\begin{figure}[H]
    \centering 
	\includegraphics[width=0.48\textwidth]{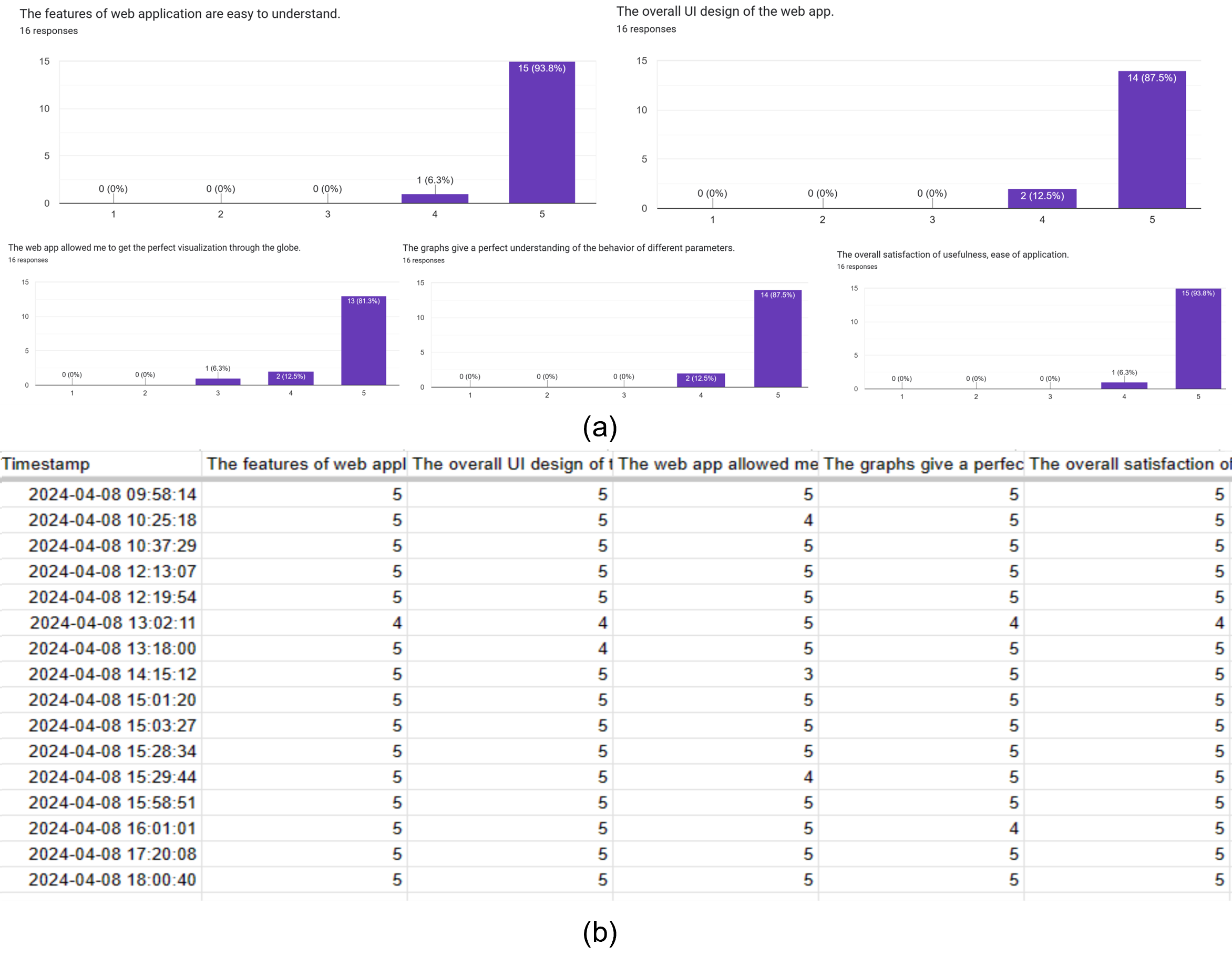}
	\caption{(a) Validation Survey Result (b) Validation Result with Timestamp generated by Google form in CSV format}
	\label{fig::13}
\end{figure}

Figure \ref{fig::13} provides evidence that the usability study resulted in mainly positive outcomes, demonstrating user contentment with the application's consistency, user-friendliness, ease of learning, and efficiency.

\section{My Contribution to this Project}

Each team member took ownership of preprocessing and visualizing a specific dataset. My responsibility centered on the Global Terrorism dataset \cite{r4}, where I constructed a visualization webpage, subsequently deploying it. Furthermore, I took charge of building and deploying the homepage, seamlessly integrating all three visualization pages to create a unified platform, and creating a Google form for validation survey. 

\section{Conclusion \& Future Work}
Our project uses advanced visualization tools like Plotly, Plotly.js, and D3.js to analyze global trends in terrorism, air quality, and population dynamics. Through interactive maps and charts, we provide insights into spatial dynamics, health risks, and demographic transitions. Deployed on accessible web platforms, our visualizations democratize access to information, fostering collaboration and informed decision-making. Future plans include enhancing interactivity, integrating machine learning, and expanding datasets for improved usability and inclusivity.

\bibliography{conference_101719}{}

\begin{thebibliography}{1}

\bibitem{r5}
{W}orld {A}ir {Q}uality {I}ndex by {C}ity and {C}oordinates --- kaggle.com.
\newblock \url{https://www.kaggle.com/datasets/adityaramachandran27/world-air-quality-index-by-city-and-coordinates}, 2023.

\bibitem{r4}
{G}lobal {T}errorism {D}atabase --- start.umd.edu.
\newblock \url{https://www.start.umd.edu/gtd/}, 2024.

\bibitem{r1}
Diansheng Guo, Ke~Liao, and Michael Morgan.
\newblock Visualizing patterns in a global terrorism incident database.
\newblock {\em Environment and Planning B: Planning and Design}, 34(5):767--784, 2007.

\bibitem{r2}
Xinchi Luo, Runfeng Jiang, Bin Yang, Hongxing Qin, and Haibo Hu.
\newblock Air quality visualization analysis based on multivariate time series data feature extraction.
\newblock {\em Journal of Visualization}, pages 1--18, 2024.

\bibitem{r6}
United Nations.
\newblock World population prospects - population division.
\newblock \url{https://population.un.org/wpp/}, 2022.

\bibitem{r3}
Max Roser and Hannah Ritchie.
\newblock Two centuries of rapid global population growth will come to an end.
\newblock {\em Our World in Data}, 2024.

\end{thebibliography}
\bibliographystyle{plain}

\end{document}